# Modelling the coupling between ocean and atmosphere; the global signature of the El Niño/La Niña Southern Oscillation

Costas A. Varotsos[1] and Arthur P. Cracknell[2]

Climate Research Group, Division of Environmental Physics and Meteorology, Faculty of Physics, National and Kapodistrian University of Athens, University Campus Bldg. Phys. V, Athens 15784, GR

[2]Division of Electronic Engineering and Physics, University of Dundee, Dundee, DD1 4HN, UK

**Abstract**

Satellite and ground-based observations are used to explore the composite oceanic - atmospheric link known as the El Niño/La Niña Southern Oscillation (ENSO) phenomenon, which is closely associated with extreme weather events (e.g. heat waves, tornadoes, floods and droughts), incidence of epidemic diseases (e.g. malaria), severe coral bleaching, etc. The ENSO temporal evolution depends on the energy exchange between the coupled ocean/atmosphere system. Its cycle has an average period of about four years, but there is considerable modulation of it from several sources and this is not yet fully understood. This paper aims to give a better insight to the global signature of ENSO evolution considering both the continuous natural interactions taking place between ocean and atmosphere, and anthropogenic effects. The results obtained could be employed to elucidate the development of more accurate advanced modelling of ocean - atmosphere interactions, thereby improving climate change projections.

## 1. Introduction

### 1.1 Definitions

Traditionally, the term El Niño / La Niña refers to the process of the irregular occurrence of the oceanic event of extensive heating / cooling of the central and eastern tropical Pacific. El Niño / La Niña has a significant impact on the shift of weather systems along the Pacific, such as increased convection (e.g. changing Walker circulation), altering cloudiness in the central tropical Pacific Ocean and weaker or stronger than normal trade winds in the Pacific Ocean.

A well-established link between the atmosphere and the oceanic El Niño/La Niña phenomenon is the seesaw back and forth in surface air pressure between the eastern and the western South Pacific, which is known as Southern Oscillation (SO).
The strength of this oscillating atmospheric phenomenon is measured by the Southern Oscillation Index (SOI), which is computed from the difference in the monthly surface air pressure between Tahiti (17° 40′S, 149° 25′W) and Darwin (12° 27′S, 130° 50′E). It is widely recognized that El Niño episodes are associated with negative values of the SOI, while La Niña episodes are associated with positive SOI values.

There is a composite oceanic-atmospheric phenomenon known as the El Niño/La Niña Southern Oscillation (ENSO) phenomenon, which is a quasi - periodic phenomenon (with a typical periodicity of 3–7 years) resulting from the interaction of the ocean and the atmosphere with many impacts on climatic and weather conditions not only in the tropical Pacific but in many regions of the world[1]. Tudhope et al.[2] using annually banded corals from Papua New Guinea showed that ENSO has existed for the past 130,000 years, operating even during "glacial" times of substantially reduced regional and global temperature and changed solar forcing. Tudhope et al.[2] also found that during the 20th century ENSO has been strong compared with ENSO of previous cool (glacial) and warm (interglacial) times. They suggested that the observed pattern of change in amplitude may be due to the combined effects of ENSO dampening during cool glacial conditions and ENSO forcing by precessional orbital variations.

*1.2 The contribution of remote sensing to ENSO- related research*

As mentioned above ENSO activity can profoundly impact several climate ecosystems. As far as the marine ecosystem is concerned during the 1997–1999 El Niño/La Niña transition period, phytoplankton biomass increased by 10 % globally[3,4]. Short-term variability (less than one decade) in chlorophyll concentration, primary production and phenology of phytoplankton populations have been shown to correlate with ENSO variability in the Equatorial regions and in the global oceans.

The remote sensing techniques offered unique observations for the investigation of those ecosystems with high temporal and spatial coverage. For example, ocean-colour sensors on satellites can provide estimates of chlorophyll concentration at high spatial and temporal resolutions and at global scale. Because they provide data consistently and frequently and over long periods of time, they are suitable for computations of several ecological indicators and for studying inter-annual variations and long-term trends in the state of the marine ecosystem. However, ocean-colour sensors do have a finite lifespan, and differences in instrument design and algorithms make it difficult to compare data from multiple sensors (Racault et al. 2017). When overlapping data are available from two or more sensors, such data can be used to establish inter-sensor bias and correct for it. The ESA Ocean Colour CCI (OC-CCI) has merged ocean-colour data using the Sea-viewing Wide Field-of-View Sensor (SeaWiFS 1997–2010), the Moderate-Resolution Imaging Spectroradiometer (MODIS 2002-present) and the MEdium-Resolution Imaging Spectrometer (MERIS 2002–2012) to provide the first 17-year (1997 to present) global scale, high-quality, bias-corrected and error-characterised data record of ocean colour[5]. In addition, the strong 2015-16 ENSO event coincided with unprecedented coverage of space-based remote sensing of sea surface salinity over global oceans, notably: from the NASA's missions of Soil Moisture Active Passive (SMAP) and Aquarius, and the ESA's Soil Moisture and Ocean Salinity (SMOS)[6.]

Another example is the use of data acquired in the middle infrared wavelengths by sensors such as the NOAA AVHRR, Terra MODIS and Envisat Advanced Along Track Scanning Radiometer, which are strongly moisture sensitive, provide vegetation index data that are strongly related to variables associated with drought impacts.

*1.3 The concern for credible ENSO prediction*

The global concern for the prediction of ENSO stems from the fact that an ENSO is a dominant driving force of climate variability having climatological impacts in regions far away from the tropical Pacific (i.e. teleconnections) and may be linked to extreme weather conditions (e.g. floods and droughts), changes in the incidence of epidemic diseases (e.g. malaria), severe coral bleaching, civil conflicts, etc.

The devastating consequences of ENSO's strong events have endeavored for their credible short-term and long-term prediction. In this context, a new method was recently presented for the detection of precursors strong El Niño events with the use of entropy change in a new time domain termed as "natural-time" [7,8]. The analysis of the SOI time series using this modern method has given considerable early signs of two of El Niño's most powerful events (1982-1983 and 1997-1998).

Recently, an investigation was made about the forecast by the Australian Government's Meteorological Office on September 1, 2015 according to which, the most international climatic models showed that the 2015 El Niño would be the strongest since 1997-98 and it would have reached its maximum at the end of 2015 (http://www.bom.gov.au/climate/enso/archive/ensowrap_20150901.pdf). However analysing the SOI time series for the period 1876–2015 showed that the running 2015–2016 El Niño would be rather a "moderate to strong" or even a "strong" event and not "one of the strongest on record", as that of 1997–1998[7]. In fact, the 2015–2016 El Niño was finally a strong event, but not the strongest on record. This is an example of the necessity for the feasibility of accurate forecast of such crucial natural events, as El Niño. It is of great importance given that these events might be associated with the fact that the global annual average of temperature reached in 2015 the warmest values on record.

The purpose of this paper is to detect possible reasons for the significant discrepancies between ENSO's observations and modelled data. which play an important role in the weaknesses of the prediction models of extreme ENSO events. A plausible reason could be the fact that the temporal evolution of ENSO as derived from real observations exhibit particular features that are missing from the temporal evolution of ENSO derived from models. One reasonable reason could be the fact that the evolution of the ENSO as it results from the actual observations has particular characteristics but which are missing from the evolution of the ENSO as it results from the models.

## 2. Data and Analysis

The SOI data set used in this paper consists of 798 data points and represents a monthly averaged normalized difference between the pressure measured at Darwin and the sea level pressure measured at Tahiti during the period 1/1951- 6/2017. This data set was obtained from National Centers for Environmental Prediction (NCEP) - NOAA (http://www.cpc.ncep.noaa.gov/data/indices/soi) (Fig. 1a).

Another, more extensive, set of data used in the present analysis is the mean monthly values of Best ENSO Index (BEI) for the period from 01/1870 to 06/2017 (1770 points), obtained from NOAA Earth System Research Laboratory. The BEI was designed to be simple to calculate and provide a long-term ENSO index for research purposes. The BEI time series is based on the combination of an atmospheric element of the ENSO phenomenon (the SOI) and an oceanic component (the Nino 3.4 sea surface temperature - SST, defined as the SST averaged over the region 5N-5S and 170W to 120W). The monthly mean climatology for the period 1898-2000 was removed for all data. After that, the values were standardized per month so that each month it has a mean of 0 and a standard deviation of 1.0 for all years during the output time period. The resulting SST and SOI values were averaged for each month of the time series (https://www.esrl.noaa.gov/psd/people/cathy.smith/best/enso.ts.1mn.txt) (Fig. 1b).

The seasonality of the SOI and BEI time series was removed using the classical Wiener method [9,10].

In order to investigate the existence of scaling dynamics in both time series, we used the well-known detrended fluctuation analysis (DFA), which eliminates the noise of non-stationary time series and detects their scaling features[11-15].

The DFA technique provides a relationship between the root mean square fluctuations $F_d(\tau)$ and the segment size $\tau$, characterized for a power-law $F_d(\tau) \sim \tau^\alpha$. An exponent $\alpha \neq 0.5$ in a certain range of $\tau$ values indicates the existence of long-range correlations in this time interval, while $\alpha = 0.5$ corresponds to the classical random walk (white noise). When $0 < \alpha < 0.5$, power-law anticorrelations are present (antipersistence). If $0.5 < \alpha < 1.0$, then persistent long-range power-law correlations prevail; the case $\alpha = 1$ corresponds to the so-called $1/f$ noise. Moreover, if $1 < \alpha < 1.5$, then there are again long-range correlations.

It is noteworthy that crucial environmental problems exhibit non-linear dynamics and therefore their study must be implemented by employing modern statistical tools[16-19].

In the following, to look at the singularity spectrum of SOI time series and to estimate the degree of multifractality, the Multifractal Detrended Fluctuation Analysis (MF-DFA2) is used. However, the existence of long-range correlations in both the SOI and BEI time series was established detecting the type of the autocorrelation function and the stability of the local slopes at long time scales.

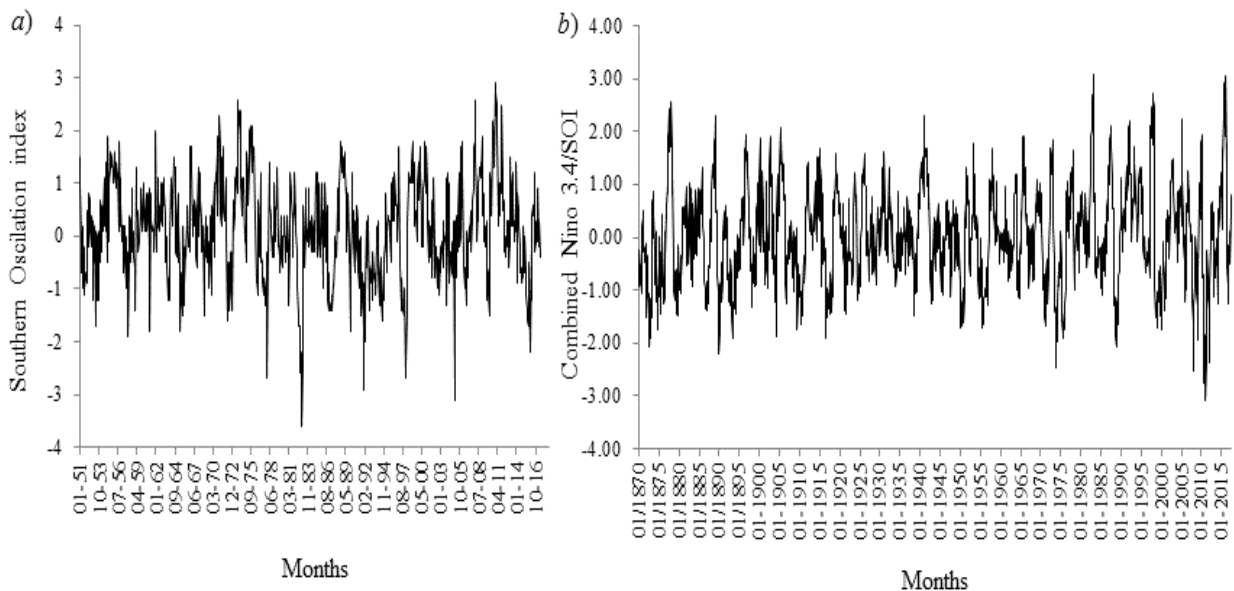

**Figure 1.** (*a*) The SOI mean monthly values for the period from January 1951 to June 2017 representing the standardized pressure difference between Tahiti and Darwin stations (798 data points). (*b*) The BEI mean monthly values during 01/1870 - 06/2017 combining two components of the ENSO phenomenon: the SOI and the Nino 3.4 SST (1770 data points).

## 3. Discussion and Results

### *3.1 Power law distribution in the extreme SOI fluctuations*

Our first step was to calculate the $y_{i+1} - y_i$ temporal increments of the SOI mean monthly values ($y_i$) representing the standardized pressure difference between Tahiti and Darwin stations for the period from January 1951 to June 2017.

These increments were grouped into classes of equal length, in order to examine the hypothesis whether SOI fluctuations obey the Gaussian distribution. However, the statistical best-fit tests of Kolmogorov–Smirnov, chi-square and Anderson–Darling, led to the rejection of the above mentioned hypothesis (at a 95% confidence level) and extreme values (defined as those that are located out of the $[m - 2\sigma, \; m + 2\sigma]$ interval, where $m$ and $\sigma$ stand for the data set average and standard deviation, respectively) seemed to be mainly responsible for the poor representativeness of the normal distribution for SOI fluctuations (Fig. 2a).

To determine the distribution that best fits the SOI increments data set, the empirical probability $P(X > x)$ of exceeding a fixed SOI increment of amplitude $x = |y_{i+1} - y_i|$ was calculated and then plotted in a semi-logarithmic graph, against the value of $x$ (Fig. 2b). The tail of this distribution seemed to be consistent with the power law $P(X > x) \sim x^{-\mu}$, indicating a decrease of the probability of increasing the intensity of the time increments. A linear least-squares fit revealed an exponent $\mu = 4.52 \pm 0.17$ (with coefficient of determination $R^2 = 0.98$) when the amplitudes of increments were between 1.3 and 2.6 and the value 1.3 coincided with the value $m + 2\sigma$. It is worth noting that the value of $\mu = 4.52$ is well out of the range for stable Levy distributions, $0 < \mu < 2$ and especially outside the Gaussian distribution ($\mu = 2$), indicating that extreme fluctuations are more likely to occur than the Gaussian distribution would predict.

For validation purposes we re-apply the above analysis to another, more extensive data set comprised of the mean monthly values of BEI (i.e. the combined SOI and Nino 3.4 SST) for the period from 01/1870 to 06/2017 (1770 points). The results of this analysis are presented in Table 1.

**Table 1.** Results of the analysis applied on the temporal increments ($y_{i+1} - y_i$) of BEI mean monthly values ($y_i$)

| | |
|---|---|
| **Goodness-of-fit tests for the normal distribution, on BEI increments $y_{i+1} - y_i$:** | The statistical best-fit tests of Kolmogorov–Smirnov, chi-square and Anderson–Darling, suggested rejection of Gaussian distribution fit (at 95% confidence level). |
| **Log-log plot of the empirical probability $P(X > x)$, where $x$ stands for $\|y_{i+1} - y_i\|$ :** | The linear least-squares fit ($y = -5.75x - 1.19$ with coefficient of determination: $R^2 = 0.97$) on the log-log plot of $P(X > x)$ vs. a fixed BEI increment $x$ gave an exponent $\mu = 5.75 \pm 0.14$ when the amplitudes of increments were between 0.99 and 2.09 and the value 0.99 coincided with the value $m + 2\sigma$. |

**Notice:** The value $\mu = 5.75$ is out of range for stable Levy distributions, indicating again that extreme BEI fluctuations are more likely to occur than would be expected from the Gaussian distribution.

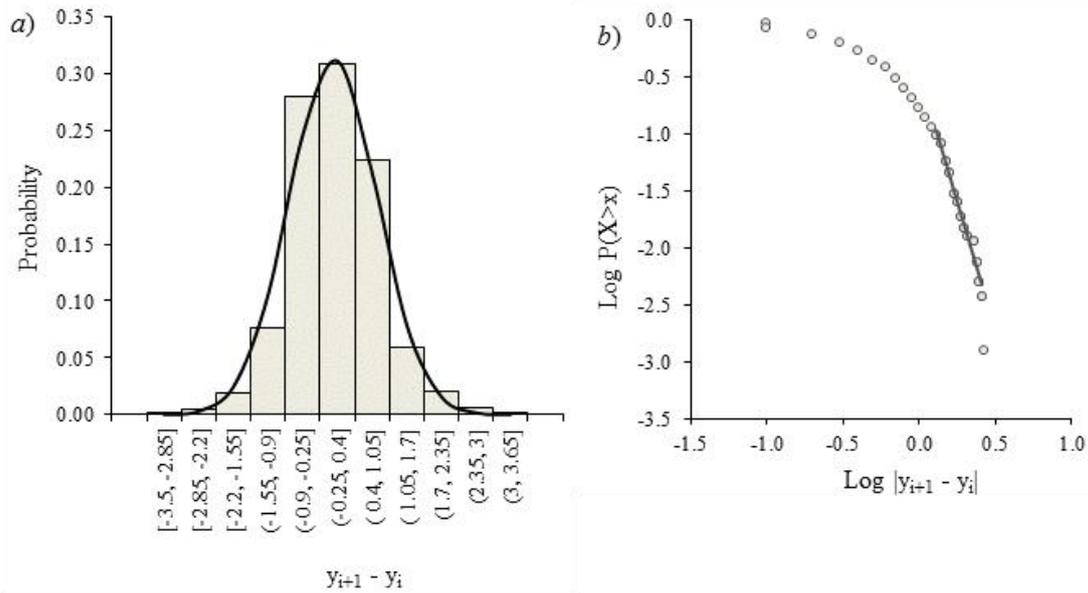

**Figure 2.** (*a*) The histogram of the SOI increments data set. The smooth line shows the normal distribution. (*b*) Log-log plot of empirical probability to observe SOI increments with amplitude greater than some value $x = |y_{i+1} - y_i|$. The linear least-squares fit ($y = -4.52x - 0.45$, with $R^2 = 0.98$) shows an asymptotic power law scaling for fluctuations amplitudes between 1.3 and 2.6, where 1.3 coincides with the value $m + 2\sigma$.

*3.2 Power law in the SOI temporal evolution*

The behaviour of the asymptotic power law of the SOI temporal increments distribution, mentioned in the previous section, seemed to be a particularly suitable description of the occurrence of extreme events. So, our next step was to study the scaling features of SOI time series using the DFA technique.

Figure 3(a) illustrates the log–log plot of the root mean square fluctuation function $F_d(\tau)$, for the SOI time series, versus temporal interval $\tau$ (in months). Focusing on this figure, it is clear that DFA application in the SOI time series resulted in a crossover at approximately 2.5 years (i.e. $\log\tau = 1.46$) while the corresponding value of the root mean square fluctuation function $F_d(\tau)$ was 2.6 (i.e. $\log F_d(\tau) = 0.42$). The slope of the corresponding best fit equation ($a = 1.12 \pm 0.02$) before the crossover point reveals the persistent behaviour (of $1/f$ - type) for the SOI temporal evolution. This suggests that SOI fluctuations at short time intervals are associated with SOI fluctuations at longer time intervals in a power-law mode (and the corresponding $F_d(\tau)$ values are less than or equal to 2.6 for time scales $\leq 2.5$ years). On the other hand, according to the asymptotic power law scaling detected in the plot of the empirical probability $P(X > x)$ vs a fixed SOI fluctuation $x$, the SOI extreme fluctuations (with amplitudes from 1.3 to 2.6) appear to be more likely to happen than the Gaussian distribution would predict and the probability of their re-appearance fits to power law distribution.

However, after the detected crossover point the scaling properties of the SOI time series seem to change. The slope of the corresponding best fit equation ($a = 0.65 \pm 0.03$) shows persistent behaviour but with weaker memory than the one preceding to crossover point.

Our next step was to establish the power-law long-range correlations in the SOI time series that we investigated the rejection of the exponential decay of the autocorrelation function and the constancy of "local slopes" in a certain range to the low frequencies. Prior to this, we removed the seasonality of the SOI time series (by using the classical Wiener method) in order to eliminate the above discussed crossover.

In more detail, we apply the DFA technique again to the deseasonalised SOI time series and the single linear fitting on the entire range of scales gave the exponent $a$ = 0.89 ± 0.02 (Fig. 3b). We then evaluated the local slopes of $\log F_d(\tau)$ vs. $\log \tau$ (separately for two different window sizes of 10 and 12 points, which were shifted successively to all calculated scales $\tau$), requesting constancy in a sufficient range. For this purpose, we performed Monte Carlo simulations by applying the DFA method in 500 times series (with 798 points) characterized by fractional Gaussian noise (with $a$ = 0.89), to calculate the local slopes-$a(\tau)$ for each of the 500 time series, in a 10-point window that was successively shifted to all calculated scales $\tau$.

Thus, we determined a range $R$ for the $a(\tau)$-local slopes of SOI time series, based on their mean value $\bar{a}$ (in all the calculated scales $\tau$), increased and decreased by the two standard deviation $\sigma_{a(\tau)}$, of the 500 estimated local slopes-$a(\tau)$, i.e. $R = (\bar{a} - 2\sigma_{a(\tau)}, \bar{a} + 2\sigma_{a(\tau)})$. According to Fig. 4(a), all the local slopes (after $\log\tau = 1.17$) lie within the boundary of the $R$ range, although they do not imply sufficient constancy.

With regard to the criterion of exponential decay of the autocorrelation function, Fig. 4(b) shows the power spectral density profile for the SOI time series, indicating that the power-law fit gives a higher coefficient of determination compared to the exponential fit. Thus, both criteria of Maraun et al. (2004) do not appear to be rejected and therefore the long-range correlations of power-law type for the SOI time series are established.

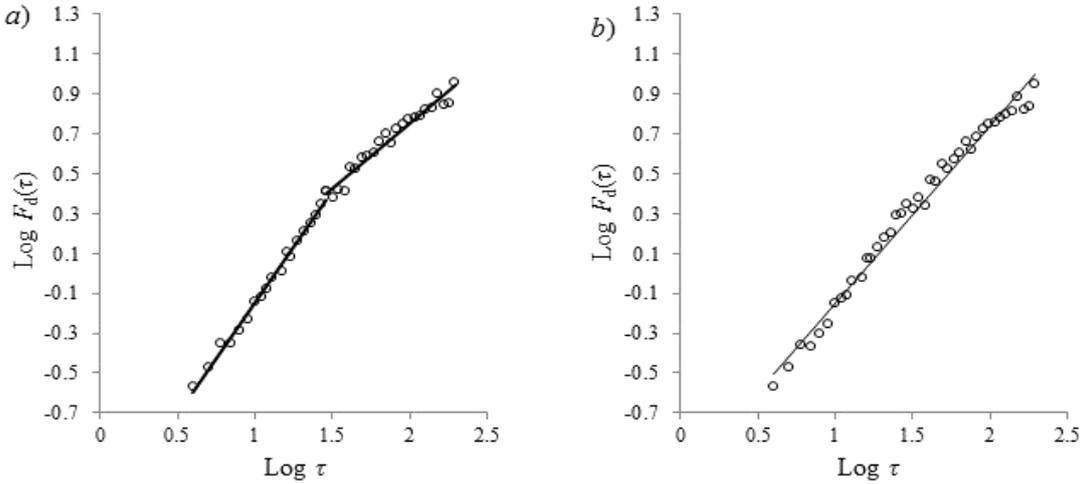

**Fig. 3.** Log-log plot of the root-mean-square fluctuation function $F_d(\tau)$ versus time scale $\tau$ (in months), (*a*) for the initial SOI time series, with the respective best fit equations (prior to crossover point: $y = 1.12x - 1.27$, $R^2 = 0.99$ and after the crossover point: $y = 0.65x - 0.54$, $R^2 = 0.96$) and (*b*) for the deseasonalised SOI time series with the corresponding best fit equation ($y = 0.89x - 1.04$, $R^2 = 0.98$).

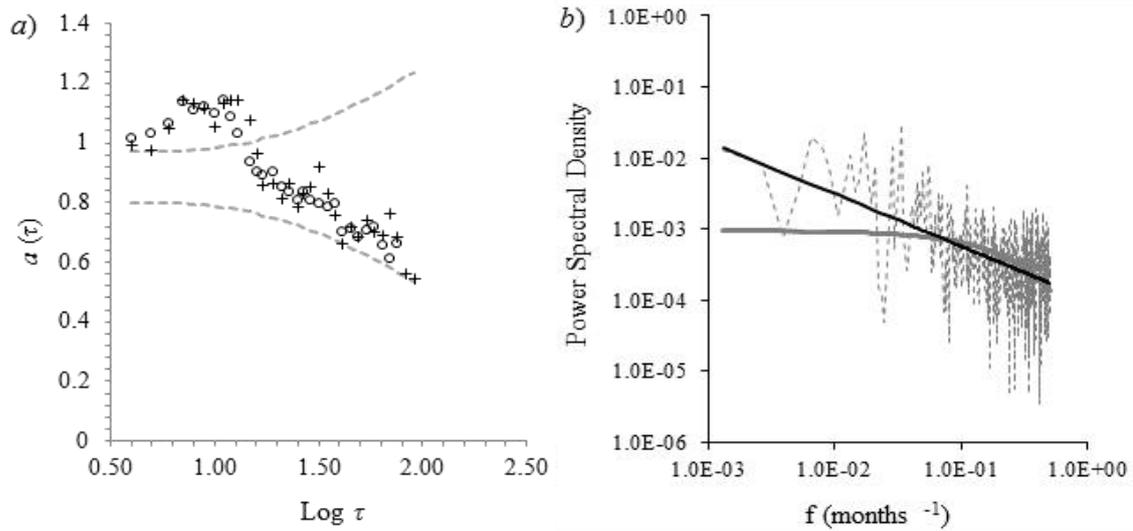

**Fig. 4.** (*a*) Local slopes of the log$F_d(\tau)$ vs. log$\tau$ calculated within a window of 10 points (crosses +) and of 12 points (circles o) for the deseasonalised SOI time series. The dashed grey line indicates the corresponding $2\sigma$ intervals around the mean value of the local slopes ($\bar{a}$ = 0.88). (*b*) Power spectral density for the deseasonalised SOI time series with the corresponding power-law (black line) and the exponential (grey line) fit ($y = 1 \cdot 10^{-4} x^{-0.74}$ with $R^2 = 0.26$ and $y = 0.0009 e^{-3.86x}$ with $R^2 = 0.16$, respectively).

However, in order to confirm all the above findings we again applied the same analysis to the BEI's time series. The results of this analysis are depicted in Table 2.

**Table 2.** Results of the analysis applied on the initial and the deseasonalised BEI time series

| | |
|---|---|
| **Log-log plot of root-mean-square fluctuation function $F_d(\tau)$ vs. time scale $\tau$ (in months):** | A crossover at $\tau = 24$ with $F_d(\tau) = 2.09$ was detected for the initial BEI time series. The linear least-squares fits (before the crossover point: $y = 1.41x − 1.62$, $R^2 = 0.99$ and after the crossover point: $y = 0.65x - 0.49$, $R^2 = 0.96$) showed persistent behaviour obviously stronger before than after the crossover point. (i.e. BEI fluctuations in short time intervals are related to those at longer time intervals in a power-law fashion and the corresponding $F_d(\tau)$ values are less than or equal to 2.09 for time scales ≤ 2 years). Furthermore, the single linear fitting ($y = 0.84x - 1.09$, $R^2 = 0.98$) across the entire range of scales for the deseasonalised BEI time series gave exponent $a = 0.84 \pm 0.02$ validating the persistent memory. |
| **Establishment of the long-range correlations for the deseasonalised BEI time series** | The local slopes of log$F_d(\tau)$ vs. log$\tau$ seemed to lie within the boundary of the $R$ range (after log$\tau = 1.2$), although they did not imply sufficient constancy. Additionally, the power spectral density profile indicated that the power-law fit ($y = 8 \cdot 10^{-6} x^{-1.21}$ with $R^2 = 0.41$) gives a higher coefficient of determination compared to the exponential fit ($y = 0.0003 e^{-6.73x}$ with $R^2 = 0.28$). |

## 3.3 Multifractality in SOI time series

The next step was to study the spectrum of singularities for deseasonalised SOI time series, employing the MF-DFA2 technique and calculating the $q^{th}$ order fluctuation function $F_q(\tau)$ for various moments $q$. According to Fig. 5 (a), the scaling behaviour of $F_q(\tau)$ (i.e. slope) for all selected positive (negative) moments $q$ is almost the same, especially for $\log\tau > 2$ ($\tau > 100$). This behaviour has shown the existence of a large degree of multifractality, especially for shorter time scales ($\tau \leq 100$).

The above proposed multifractality is defined in Fig. 5(b) where the generalized Hurst exponent $h(q)$ for SOI time series varies versus $q$ values (i.e. $h(q)$ is not independent of $q$), while the $h(q)$ values, which are obviously higher than 0.5 show long-term persistence over the time series considered. In addition, the slope of $h(q)$ for positive moments seems to be similar to that one of negative moments, which is consistent with the findings of Fig. 5(a).

Finally, Fig. 5(c) presents the singularity spectrum $f(n)$ as a function of the singularity strength $n$ for SOI time series. The maximum value of $f(n)$ corresponds approximately to q=0, while the $f(n)$ values on the left (right) of the maximum value correspond to positive (negative) moments $q$. It is obvious that $f(n)$ varies similarly to both sides of its maximum value. This again reveals common features of multifractality for positive and negative $q$-values.

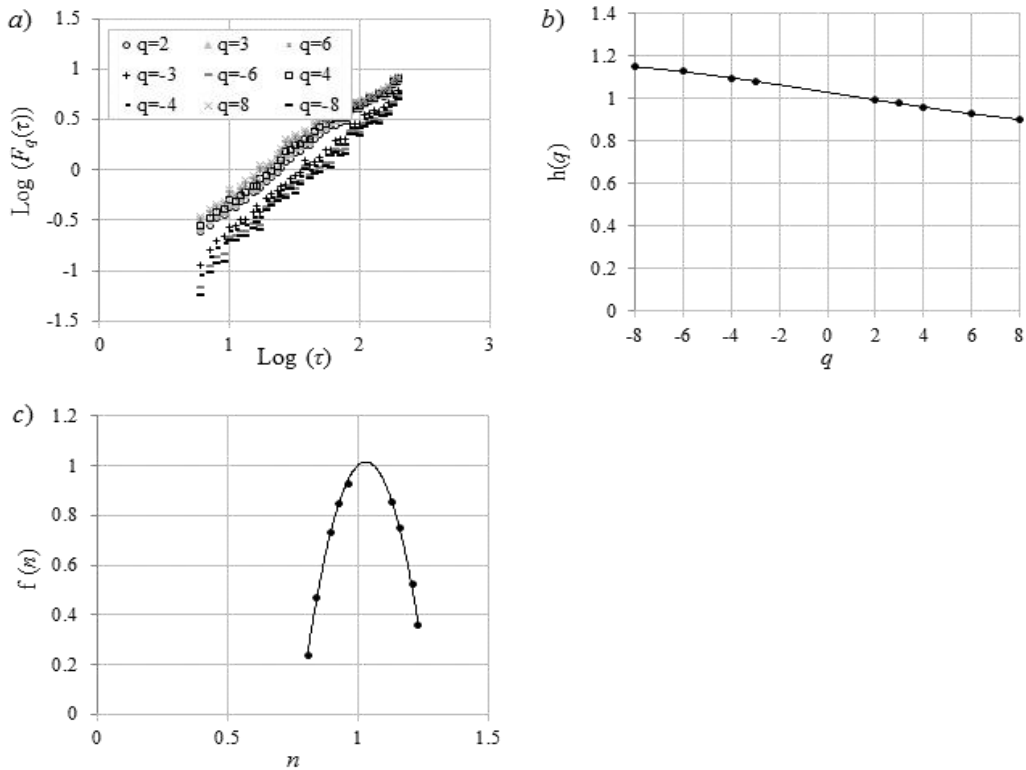

**Fig. 5.** (*a*) Log-log plots of the MF-DFA2 fluctuation factor $F_q(\tau)$ versus the time scale $\tau$ for specific moments $q$ for deseasonalised SOI time series (upper left panel). Generalized Hurst exponent $h(q)$ versus $q$ for the time series studied. (*a*) The empirical curve (dots) is fitted by the third order polynomial ($y = 4 \cdot 10^{-5}x^3 - 6 \cdot 10^{-5}x^2 - 0.018x + 1.03$, with $R^2 = 0.999$). (*c*) Singularity spectrum $f(n)$ versus singularity strength $n$ for the time series studied. The empirical curve (dots) is fitted by the third order polynomial ($y = -1.3x^3 - 11.8x^2 + 28.3x - 14.3$, with $R^2 = 0.998$).

Similar results were obtained by applying the MF-DFA2 to deseasonalised BEI time series. Multifractality was revealed for positive and negative $q$-values and $h(q)$-values, which were again higher than 0.5 indicated long-term persistence for the time series examined (see Table 3).

Table 3. MF-DFA2 technique results for the deseasonalised BEI time series

| | |
|---|---|
| **Log-log plots of the MF-DFA2 fluctuation factor $F_q(\tau)$ vs. the time scale $\tau$ for specific moments $q$:** | The scaling behaviour of $F_q(\tau)$ (i.e. slope) for all the selected positive (negative) moments $q$ was almost the same, especially for $\log\tau > 2$ ($\tau > 100$). |
| **Generalized Hurst exponent $h(q)$ vs. $q$:** | The generalized Hurst exponent $h(q)$ was not independent of $q$, while the $h(q)$ values were obviously higher than 0.5 showing long-term persistence over the time series considered ($h(q) = 7 \cdot 10^{-5}q^3 + 5 \cdot 10^{-4}q^2 - 0.023q + 0.98$, with $R^2 = 0.999$). |
| **Singularity spectrum $f(n)$ vs. singularity strength $n$:** | The singularity spectrum $f(n)$ vs. the singularity strength $n$ indicated that the maximum value of $f(n)$ corresponds approximately to $q=0$, while the $f(n)$ values on the left (right) of the maximum value correspond to positive (negative) moments $q$ ($f(n) = 2.5n^3 - 19.6n^2 + 31.2n - 13.1$, with $R^2 = 0.986$). |

**4. Conclusions**

We investigated the distribution of both SOI and BEI temporal increments (for the period 1951-2017 and 1870-2017, respectively) and the intrinsic scaling properties of the time series of the same parameters using DFA and MF-DFA2 tools. The main findings were:

1) An asymptotic power law scaling that was detected in the plot of the empirical probability $P(X > x)$ vs a fixed SOI fluctuation $x$, suggested that extreme SOI fluctuations (with amplitudes ranging from 1.3 to 2.6) appear to be more likely to happen than the Gaussian distribution would predict and the probability of their re-appearance fits within the power law distribution.
2) The DFA application in the SOI time series resulted in a crossover of about 2.5 years, while the corresponding value of the root mean square fluctuation function $F_d(\tau)$ was 2.6. The slope of the corresponding best fit equation ($a = 1.12 \pm 0.02$) before the crossover point revealed persistent behaviour (of $1/f$ - type) for the SOI temporal evolution. This suggests that SOI fluctuations at short time intervals are associated with SOI fluctuations at longer time intervals in a power-law fashion (and the corresponding $F_d(\tau)$ values are less than or equal to 2.6 for time scales ≤ 2.5 years).
3) The asymptotic power law scaling detected in the empirical probability $P(X > x)$ vs a fixed BEI fluctuation $x$ indicated that BEI extreme fluctuations (with amplitudes from 0.99 to 2.09) seemed more likely to occur than that the Gaussian distribution predicted and their chance of their reappearance fits into power law distribution.

4) The DFA application in the BEI time series led to a crossover of about 2 years while the corresponding value of the root mean square fluctuation function $F_d(\tau)$ was 2.09. The slope of the corresponding best fit equation ($a = 1.41 \pm 0.02$) before the crossover point revealed strong persistent behaviour for the BEI temporal evolution. This suggests that BEI fluctuations at short time intervals are related to BEI fluctuations at longer time intervals in a power-law fashion (and the corresponding $F_d(\tau)$ values are less than or equal to 2.09 for time scales $\leq 2$ years).
5) The deseasonalised SOI and BEI time series exhibited positive long-range correlations and multifractal behaviour.
6) The application of MF-DFA2 technique to the deseasonalised time series of SOI and BEI showed multifractality for positive and negative q-values and long-term persistence was again confirmed.

The results herewith may enhance the reliability of the models for predictions of SOI fluctuations, but also to the study of other atmospheric parameters that are closely related to climate change, such as the dynamics of the ozone layer [19-24]. Nowadays, there are long time series of the atmospheric ozone from the Earth's surface up to the mesosphere[25-32] and therefore application of similar analysis to these data will provide missing information about the temporal and spatial evolution of atmospheric ozone and its impacts to the environment, such as the air pollution problems [33-40].

**Note:** A preliminary version of this work has been submitted for presentation to the Remote Sensing & Photogrammetry Society Annual Conference: Earth & Planets: Making the most of our observations 5-8 September 2017

**Competing Interests** The authors declare that they have no competing financial interests.

**Correspondence** Correspondence and requests should be addressed C. Varotsos (email: covar@phys.uoa.gr)